\documentclass[prd,preprint,showpacs,showkeys,nofootinbib,superscriptaddress]{revtex4}

\usepackage{amsmath,amssymb}  
\usepackage{bm}               
\usepackage{graphicx}         
\usepackage{epsfig}          

\begin{document}


\preprint{BNL-NT-07/15} \preprint{RBRC-665}


\title{Asymmetric Di-jet Production in Polarized Hadronic Collisions}

\author{Jian-Wei Qiu}
\affiliation{Department of Physics and Astronomy,
             Iowa State University, Ames, IA 50011}
\affiliation{Physics Department, Brookhaven National Laboratory,
             Upton, NY 11973}
\author{Werner Vogelsang}
\affiliation{Physics Department, Brookhaven National Laboratory,
             Upton, NY 11973}
\author{Feng Yuan}
\affiliation{RIKEN BNL Research Center, Building 510A,
             Brookhaven National Laboratory, Upton, NY 11973}

\date{\today}

\begin{abstract}
Using the collinear QCD factorization approach, we study
the single-transverse-spin dependent cross section
$\Delta\sigma(S_\perp)$ for the hadronic production of two
jets of momenta $P_1=P+q/2$ and $P_2=-P+q/2$. We consider
the kinematic region where the transverse components of the momentum
vectors satisfy $P_\perp \gg q_\perp \gg \Lambda_{\rm QCD}$. For
the case of initial-state gluon radiation,
we show that at the leading power in $q_\perp/P_{\perp}$ and at
the lowest non-trivial perturbative order,
the dependence of $\Delta\sigma(S_\perp)$ on $q_\perp$ decouples
from that on $P_\perp$, so that the cross section can be factorized
into a hard part that is a function only of the single scale $P_\perp$,
and into perturbatively generated transverse-momentum dependent (TMD)
parton distributions with transverse momenta $k_\perp = {\cal O}(q_\perp)$.
\end{abstract}



\maketitle


{\bf 1. Introduction.}
Single-transverse-spin asymmetries (SSAs) in high-energy
hadronic reactions with one transversely polarized hadron were
first observed more than three decades ago \cite{Bunce}. The SSA
is defined as $A_N\equiv (\sigma(S_\perp)-\sigma(-S_\perp)) /
           (\sigma(S_\perp)+\sigma(-S_\perp))$,
the ratio of the difference and the sum of (differential)
cross sections when the hadron's spin vector, $S_\perp$, is flipped.
Recent experimental measurements of SSAs both in polarized hadronic
collisions \cite{E704, rhic} and in
semi-inclusive lepton-nucleon deep inelastic scattering
(SIDIS) \cite{hermes} have renewed the interest in
investigating the origin of SSAs in
Quantum Chromodynamics (QCD) \cite{review}.

It is believed that some SSAs are a consequence of the partons'
transverse motion inside the polarized hadron. The momentum scale of
this transverse motion is a typical hadronic scale, $\langle k_\perp
\rangle \sim$~1/fm~$\sim\Lambda_{\rm QCD}$. For observables with
only {\it one} hard scale $Q\gg \Lambda_{\rm QCD}$, the SSA should be
proportional to $\langle k_\perp \rangle/Q$ \cite{et,qs-ssa}. Such
observables only probe an averaged effect of the partons' transverse
motion. However, for observables characterized by more than one physical
scale, SSAs may directly probe the partons' transverse motion. For example,
in the case of Drell-Yan hadronic production of a lepton pair of
large invariant mass $Q$ and transverse momentum $q_\perp\ll Q$,
the pair probes the (anti-) quark's transverse motion at the
scale $q_\perp$, while the invariant mass $Q$ of the pair sets
the hard scale of the collision \cite{JiQiuVogYua06}. In this
letter, we study the SSA in hadronic production of two jets:
$A(P_A,S_\perp)+B(P_B)\rightarrow J_1(P_1)+J_2(P_2)+X$, with the jet
momenta $P_1\equiv P+q/2$ and $P_2\equiv -P+q/2$
\cite{BoeVog03,{mulders},Bomhof:2007su}. Unlike in the Drell-Yan
process or in SIDIS, the SSA in di-jet production can be generated
by both initial- and final-state interactions. We emphasize
that measurements of the SSA for di-jet production have begun at RHIC
\cite{star-dijet1}, complementing the measurements in SIDIS.

We are interested here in deriving a QCD formalism for the single
transverse-spin dependent cross section,
$\Delta\sigma(S_\perp)=(\sigma(S_\perp)-\sigma(-S_\perp))/2$, that
is valid in the kinematic region $P_\perp \gg q_\perp \gtrsim
\Lambda_{\rm QCD}$, where $P_\perp$ and $q_\perp$ are the
transverse components of the momenta $P$ and $q$, respectively, so
that the SSA provides direct information on the partons'
transverse motion. We first consider the region $P_\perp \gg
q_\perp \gg \Lambda_{\rm QCD}$, where both observed momentum
scales are much larger than the typical hadronic scale
$\Lambda_{\rm QCD}$. We calculate $\Delta\sigma(S_\perp)$ in terms
of the {\it collinear} QCD factorization approach, which is
expected to be valid in this region \cite{qs-ssa-fac}. In this
approach, the incoming partons are approximated to be collinear to
the corresponding initial hadrons, and the leading-order partonic
processes produce two back-to-back jets with zero momentum
imbalance.  The di-jet momentum imbalance,
$\vec{q}_\perp=\vec{P}_{1\perp}+\vec{P}_{2\perp}$, has to be
perturbatively generated by radiating an additional {\it hard}
parton.

\begin{figure}[t]
\begin{center}
\includegraphics[width=12cm]{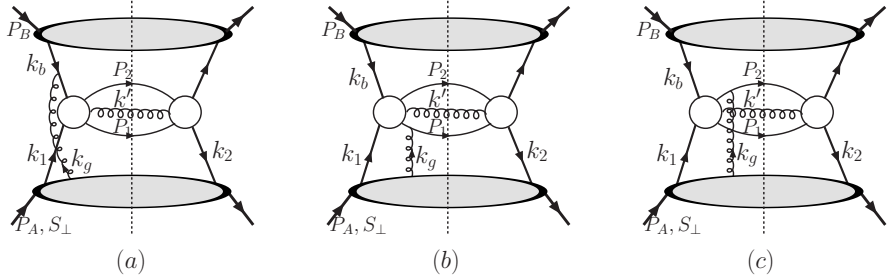}
\end{center}
\vskip -0.4cm \caption{\it Sample diagrams for quark-quark
scattering contributing to $\Delta\sigma(S_\perp)$ through an
initial-state interaction (a), and through final-state interactions with
jet $P_1$ (b) and jet $P_2$ (c).}
\label{fig1}
\end{figure}

In this letter, we concentrate on the physics issues related to
di-jet production, and restrict ourselves to the case that the
leading contribution in the expansion of the partonic scattering
in $q_\perp/P_\perp$ involves a hard $qq'\to qq'$ subprocess. We
consider the jet imbalance $q_\perp$ to be generated by gluon
radiation off the initial quarks. We derive the corresponding
leading-order contribution to $\Delta\sigma(S_\perp)$, and
demonstrate that the perturbatively calculated partonic parts can
be further factorized into a single-scale ($P_\perp$) hard part
and perturbatively generated transverse-momentum dependent (TMD)
parton distributions with transverse momenta $k_\perp = {\cal
O}(q_\perp)$. We find that the complete contributions from all
other partonic subprocesses at the leading order have the same
factorization property. These will be discussed in a forthcoming
publication~\cite{qvy-long}. The factorization of the physics at
scale $P_\perp$ from that at scale $q_\perp$ that we find when
$q_\perp \ll P_\perp$, is consistent with a more general TMD
factorization formula for the SSA in the di-jet momentum
imbalance.

{\bf 2. Single transverse-spin dependent cross section.}
When both $P_\perp$ and $q_\perp$ are much larger than
$\Lambda_{\rm QCD}$, a nonvanishing single transverse-spin dependent cross
section $\Delta\sigma(S_\perp)$ is generated by
the Efremov-Teryaev-Qiu-Sterman (ETQS) mechanism \cite{et,qs-ssa}
in the collinear factorization approach.
The calculation of $\Delta\sigma(S_\perp)$ then requires to evaluate
partonic processes with 3-parton initial- and final-states \cite{qs-ssa-fac}.
In Fig.~\ref{fig1} we show generic diagrams for the
quark-quark scattering channel that contribute to
$\Delta\sigma(S_\perp)$ through initial- and final-state interactions
with the gluon of momentum $k_g$,
which is needed for generating the phase required for
a nonvanishing SSA \cite{et,qs-ssa}. Radiation of a hard gluon of
momentum $k'$ into the final state generates the jet imbalance
$q_\perp$. The blob in the center represents tree-level
Feynman diagrams with the given initial- and final-state partons.
In the ETQS formalism, the contribution of the subprocess
$(g)qq'\to qq'g$ to $\Delta\sigma(S_\perp)$, shown in Fig.~\ref{fig1},
is generically given by
\begin{eqnarray}
&& {\hskip -0.2in}
\frac{d\Delta \sigma(S_\perp)_{(qq')}}
     {dy_1dy_2dP_\perp^2d^2\vec{q}_\perp}
=
\int \frac{dx'}{x'}\, {dx_1dx_2}\,
      T_F(x_1,x_2)\, q'(x')
\nonumber\\
&& {\hskip 0.6in}
\times
\frac{1}{16s(2\pi)^4} \,
\delta((k')^2) \,
{\cal H}_{(g)qq'\to qq'g}
\, ,
\label{x-ssa}
\end{eqnarray}
where $y_1$ and $y_2$ are the rapidities of the two jets,
$s=(P_A+P_B)^2$, and ${\cal H}$ represents a partonic hard
part. $q'(x')$ is the usual quark
distribution at momentum fraction $x'$
in the incoming hadron $B$. $x_1$ and $x_2$ are the momentum fractions
of the quarks from the polarized hadron $A$ on the two sides
of the cut shown in Fig.~\ref{fig1}, and $T_F(x_1,x_2)$ is the
corresponding twist-three quark-gluon correlation function,
extracted from the lower blob in the figure
\cite{qs-ssa,{Kouvaris:2006zy}}:
\begin{eqnarray}
T_F(x_1,x_2)
&\equiv &
\int\frac{d\zeta^-d\eta^-}{4\pi}
e^{i(x_1 P_A^+\eta^-+(x_2-x_1)P_A^+\zeta^-)}
\nonumber \\
&\times &
\epsilon_\perp^{\beta\alpha}S_{\perp\beta} \,
\left\langle P_A,S|\overline\psi(0){\cal L}(0,\zeta^-)\gamma^+
\right.
\label{TF} \\
&\times &
\left.
g{F_\alpha}^+ (\zeta^-)
{\cal L}(\zeta^-,\eta^-)
\psi(\eta^-)|P_A,S\right\rangle  \ ,
\nonumber
\end{eqnarray}
where ${\cal L}$ is the proper gauge link to make the matrix
element gauge invariant, and where the sums over color and spin indices
are implicit. In Eq.~(\ref{x-ssa}) and the rest of this
paper, the dependence on factorization and renormalization scales is
suppressed.

Equation~(\ref{x-ssa}) applies when $q_\perp\sim P_\perp$. Our
goal is now to investigate the leading structure that emerges from
Eq.~(\ref{x-ssa}) when $q_\perp\ll P_\perp$. In this limit the gluon
of momentum $k'$ is radiated either nearly collinearly from one of
the external quark legs and/or is soft. In the present work,
we only discuss collinear emission by one of the {\it initial}
quarks. This radiation is the most interesting from
the point of view of studying the factorization properties
of the cross section at small $q_\perp$, because TMD factorization
can only hold if the initial-state collinear radiation leads to a
certain specific structure, as we shall discuss below.
Since we are considering the production of jets
(as opposed to that of two specific hadrons), collinear radiation
from {\it final}-state quarks becomes part of the jet
and will not produce leading behavior in $q_\perp/P_\perp$.
On the other hand, large-angle soft gluons produced by the
interference of initial- and final state radiation may give
leading contributions, through a so-called soft factor. We leave the
detailed study of the soft factor to future work, but will briefly
return to it later.

\begin{figure}[t]
\begin{center}
\includegraphics[width=12cm]{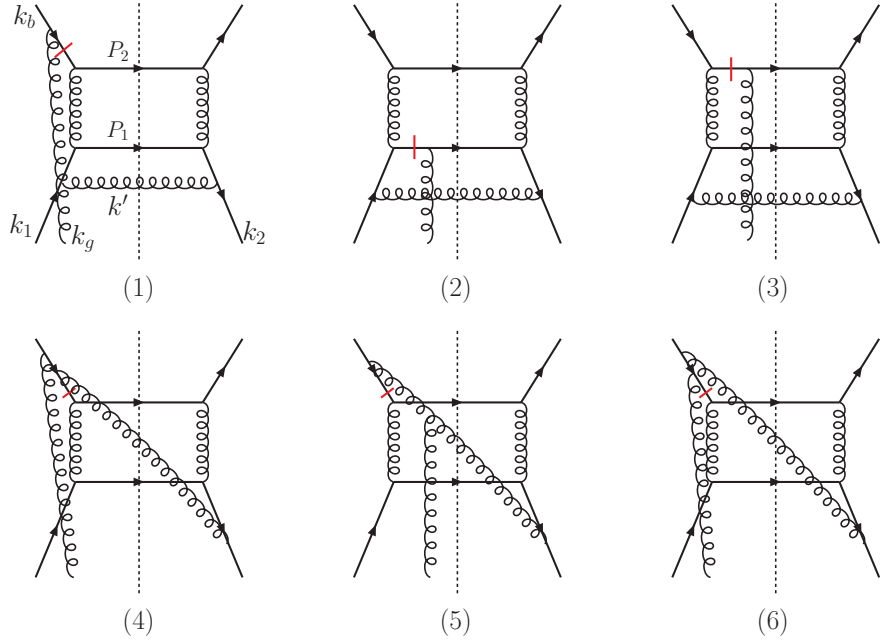}
\end{center}
\vskip -0.4cm \caption{\it Sample diagrams for the soft-pole (1-3)
and hard-pole (4-6) contributions. We have indicated the momentum
$k'$ of the radiated gluon that produces the jet imbalance $q_\perp$,
and the momentum $k_g$ of the additional gluon from the polarized
proton. The pole of each diagram is taken from the
propagator with a short bar. Diagram (1) is an example of an
initial-state interaction, and (2) and (3) show final-state
interactions. Diagrams (4-6) show the complete set of diagrams
for one hard pole, indicated again by the short bar.} \label{fig2}
\end{figure}

As we mentioned above, the strong interaction phase necessary for
a nonvanishing $\Delta\sigma(S_\perp)$ arises from the
interference between the imaginary part of the partonic scattering
amplitude with the extra polarized gluon of momentum $k_g=x_g P_A$
and the real scattering amplitude without the gluon in
Fig.~\ref{fig1}. The imaginary part comes from taking the pole of
the parton propagator associated with the integration over the
gluon momentum fraction $x_g=x_2-x_1$. For a process with two
physical scales, $P_\perp$ and $q_\perp$, tree scattering diagrams
in Fig.~\ref{fig1} have two types of poles, corresponding to
$x_g=0$ (``soft-pole'') \cite{qs-ssa} and $x_g\neq 0$
(``hard-pole'') \cite{JiQiuVogYua06}. With the extra initial-state
gluon attachment of momentum $k'$, there are many more diagrams
that contribute to $\Delta\sigma(S_\perp)$ in comparison to the
spin-averaged cross section \cite{qvy-long}. In Fig.~\ref{fig2},
we show some sample diagrams that give the leading soft-pole (1-3)
and hard pole (4-6) contributions to $\Delta\sigma(S_\perp)$ at
$q_\perp\ll P_\perp$.
By using the power counting technique \cite{css-fac}, we are able
to classify all Feynman diagrams into different groups
\cite{qvy-long}. For example, diagrams (1-3) only make leading
contributions when the momentum $k'$ of the radiated gluon is
parallel to $P_A$, whereas diagrams (4-6) can give a leading
contribution when $k'$ is either parallel to $P_A$ or to $P_B$.
Beyond that, the calculation of the soft-pole and hard-pole contributions to
$\Delta\sigma(S_\perp)$ follows the same procedure as introduced in
Ref.~\cite{JiQiuVogYua06} for the SSA in Drell-Yan and SIDIS, because
the kinematic limit considered is similar. In order to extract the
contributions to $\Delta\sigma(S_\perp)$ in Eq.~(\ref{x-ssa}), we need to
convert the extra gluon field operator in the hadronic matrix
element of the polarized hadron in Fig.~\ref{fig1} to a field strength
operator in the definition of $T_F(x_1,x_2)$ in Eq.~(\ref{TF}).
Working in Feynman gauge, we first give the initial-state
collinear partons from the polarized hadron a small transverse
momentum, $k_i = x_i P_A + k_{i\perp}$ with $i=1,2$, and then
expand the calculated partonic scattering amplitudes around
$k_{i\perp}=0$, or equivalently,
$k_{g\perp}=k_{2\perp}-k_{1\perp}=0$. The contribution to
$\Delta\sigma(S_\perp)$ arises from terms linear in $k_{g\perp}$.
After summing up all contributions \cite{qvy-long}, we obtain the
total leading-power contribution to $\Delta\sigma(S_\perp)$ from
the $(g)qq'\to qq'g$ partonic subprocess in the $q_\perp/P_\perp$
expansion:
\begin{eqnarray}
&&\left. {\hskip -0.2in}
\frac{d\Delta\sigma(S_\perp)_{(qq')}}
     {dy_1dy_2dP_\perp^2d^2 \vec{q}_\perp}
\right|_{P_\perp\gg q_\perp \gg \Lambda_{\rm QCD}}
=-\frac{\epsilon^{\alpha\beta}S_\perp^\alpha q_\perp^\beta}
     {(\vec{q}_\perp^2)^2}H_{qq'\to qq'}^{\rm Sivers}\,
\nonumber\\
&& {\hskip -0.1in}
\times
\frac{\alpha_s}{2\pi^2} x_ax_b
\left[
  q'(x_b)\int \frac{dx}{x} {\cal A}
+ T_F(x_a,x_a) \int \frac{dx'}{x'} {\cal B}
\right] ,
\label{x-ssa-qq}
\end{eqnarray}
where
\begin{eqnarray}
{\cal A}
&=&   \left\{ x\frac{\partial}{\partial x}T_F(x,x)\,
      \frac{1}{2N_c}\left[1+\xi^2\right]
\right.
\nonumber\\
&& \hskip 0.2in + T_F(x,x)\,
  \frac{1}{2N_c}\left[\frac{2\xi^3-3\xi^2-1}{1-\xi}\right]
\label{calA}\\
&& \hskip 0.2in \left. + T_F(x,\xi x)\,
  \left(\frac{1}{2N_c}+C_F\right)
  \left[\frac{1+\xi}{1-\xi}\right]
\right\} \; ,
\nonumber\\
{\cal B} &=&\, q'(x')\, C_F \left[\frac{1+\xi^{\prime
2}}{1-\xi'}\right]\ , \label{calB}
\end{eqnarray}
where $\xi=x_a/x$ and $\xi'=x_b/x'$ with
$x_a=\frac{P_\perp}{\sqrt{s}}\, (e^{y_1}+e^{y_2})$ and
$x_b=\frac{P_\perp}{\sqrt{s}}\, (e^{-y_1}+e^{-y_2})$. The above relations
are regularized at the integration limits by ``plus''-distributions
\cite{JiQiuVogYua06}. The single-scale partonic hard part is given by
\begin{equation}
H_{qq'\to qq'}^{\rm Sivers}(\hat{s},\hat{t},\hat{u})
= \frac{\alpha_s^2\pi}{\hat s^2}
\left[\frac{N_c^2-5}{4N_c^2}\right]
\frac{2(\hat s^2+\hat u^2)}{\hat t^2} \; ,
\label{hx-qq-sivers}
\end{equation}
where the use of the superscript ``Sivers'' will become clear in
the next section, and where the partonic Mandelstam variables are given
as $\hat{s}=x_a x_b s$, $\hat t=-P_\perp^2(e^{y_2-y_1}+1)$, and
$\hat u=-P_\perp^2(e^{y_1-y_2}+1)$. We note that to the leading power
in $q_\perp/P_\perp$, we have
$P_1 = P_\perp\, (e^{y_1}/\sqrt{2},
                  e^{-y_1}/\sqrt{2}, 1)$
and $P_2 = P_\perp\, (e^{y_2}/\sqrt{2},
                      e^{-y_2}/\sqrt{2}, -1)$
for the jet momenta in light-cone coordinates.

The hard part for the partonic channel we have considered,
$H_{qq'\to qq'}^{\rm Sivers}$, is very similar to
the spin-averaged partonic differential cross section
$d\hat\sigma/d\hat t$ \cite{qvy-long}. The only difference is the
color factor in square brackets. In fact, the color factor for
$H_{qq'\to qq'}^{\rm Sivers}$ is equal to a sum of three color
factors: $C_I+C_{F_1}+C_{F_2}$, corresponding to color factors of
scattering amplitudes when the initial-state gluon of momentum
$k_g$ is attached to the initial-state incoming quark of momentum
$k_b$, the final-state quark with $P_1$, or the final-state quark
with $P_2$, respectively. For the $(g)qq'\to qq'g$ subprocess, we
have \cite{qvy-long} $C_I = - \frac{1}{2N_C^2}$, $C_{F_1} = -
\frac{1}{4N_C^2}$, and $C_{F_2} = \frac{N_c^2-2}{4N_C^2}$. As a
result, the final-state interaction with the jet of momentum $P_2$
dominates, and the overall color factor $C_I+C_{F_1}+C_{F_2}$ has
a sign opposite to that of $C_I$ alone.

{\bf 3. Factorization in terms of TMD distributions.} When
$q_\perp \ll P_\perp$, the di-jet production and Drell-Yan process
at low transverse momentum ($q_T\ll Q$) considered in
Ref.~\cite{JiQiuVogYua06} share very similar kinematics. The difference is
that the Drell-Yan process has only initial-state interactions while for
di-jet production both initial- and final-state interactions are present.
It is known \cite{Collins:1984kg,JiMaYu04} that when $q_\perp\ll
Q$, the Drell-Yan cross section in leading order in $q_\perp/Q$ can be
calculated from a generalized QCD factorization formula involving the
TMD parton distributions. It is natural to ask if the generalized
QCD factorization formula can be extended to the di-jet
cross section for $q_\perp \ll P_\perp$.

When $q_\perp \gg \Lambda_{\rm QCD}$, TMD parton distributions
can be calculated in perturbative QCD (pQCD) from hard radiation and
parton splitting \cite{JiQiuVogYua06}.
The unpolarized TMD quark distribution is well known
and given by \cite{JiQiuVogYua06}
\begin{equation}
q(x_b,q_\perp)= \frac{\alpha_s}{2\pi^2}\,
\frac{1}{\vec{q}_\perp^2}\int\frac{dx'}{x'}
\left[{\cal B}+\cdots\right]\
, \label{tmd-q}
\end{equation}
where ${\cal B}$ is given in Eq.~(\ref{calB}),
and where the ellipses denote terms
proportional to $\delta(1-\xi)$ and terms generated by gluon
splitting. The transverse-spin dependent TMD quark distribution,
$q_T(x_a,q_\perp)$, known as the Sivers function \cite{Siv90}, can
also be calculated in pQCD and is given in terms of the twist-3
quark-gluon correlation function $T_F$ as \cite{JiQiuVogYua06}:
\begin{eqnarray}
q_T^{\rm SIDIS}(x_a,q_\perp) &=& -\frac{\alpha_s}{4\pi^2}
\frac{2M_P}{(\vec{q}_\perp^2)^2} \int\frac{dx}{x}
\left[{\cal A}+ \dots
\right] \ , \label{sivers}
\end{eqnarray}
where ${\cal A}$ is given in Eq.~(\ref{calA}),
$M_P$ is a hadron mass scale introduced to keep
$q(x_b,q_\perp)$ and $q_T(x_a,q_\perp)$ at the same dimension,
and where the ellipses as above denote terms proportional
to $\delta(1-\xi)$ and terms not
relevant to the following discussion.  In Eq.~(\ref{sivers}), the
superscript ``SIDIS'' indicates the Sivers function for the SIDIS
process, which has an opposite sign from that for the Drell-Yan
process given in Ref.~\cite{JiQiuVogYua06}, due to the difference
in the directions of the gauge link that defines the TMD quark
distributions \cite{BroHwaSch02,Col02,BelJiYua02,BoeMulPij03}.

One of the important features of the di-jet cross section
calculated above to leading order in $q_\perp/P_\perp$
is the separation of two observed physical scales, $P_\perp$ and
$q_\perp$. We can use Eqs.~(\ref{tmd-q}) and (\ref{sivers}) to rewrite
the single transverse-spin dependent di-jet cross section in
Eq.~(\ref{x-ssa-qq}) in terms of a single-scale
hard factor, $H^{\rm Sivers}$, which is a function of $P_\perp$, and
of the $q_\perp$-dependent perturbatively generated
TMD parton distributions:
\begin{eqnarray}
&&\left. {\hskip -0.2in}
\frac{d\Delta\sigma(S_\perp)_{(qq')}}
     {dy_1dy_2dP_\perp^2d^2\vec{q}_\perp}
\right|_{P_\perp\gg q_\perp\gg \Lambda_{\rm QCD}}
=
{\epsilon^{\alpha\beta}S_\perp^\alpha q_\perp^\beta}\,
H_{qq'\to qq'}^{\rm Sivers}
\nonumber\\
&& {\hskip 0.3in}
\times \left[\frac{1}{M_P} \left(x_b q'(x_b)\right)
       \left(x_a q_T^{\rm SIDIS}(x_a,q_\perp)\right)
\right.
\label{x-ssa-qq-tmd} \\
&& {\hskip 0.6in}
\left. -  \frac{1}{\vec{q}^2_\perp}\,
   \left(x_b\, q'(x_b,q_\perp)\right)
       \left(x_a\, T_F(x_a,x_a)\right)
\right] \, .
\nonumber
\end{eqnarray}
Using the leading order relation \cite{BoeMulPij03}
\begin{equation}
\frac{1}{M_P}\int d^2k_{\perp}\, \vec{k}^2_\perp\, q_T^{\rm
SIDIS}(x,k_\perp) = - T_F(x,x)\, ,
\end{equation}
we find that our result in Eq.~(\ref{x-ssa-qq-tmd}), when the
contributions from all other partonic subprocesses at the same
order \cite{qvy-long} are added, is consistent with the leading-order
term of a more general factorization formula
in terms of TMD parton distributions:
\begin{eqnarray}
&& {\hskip -0.1in}
\frac{d\Delta\sigma(S_\perp)}
     {dy_1dy_2dP_\perp^2d^2\vec{q}_\perp}
=
\frac{\epsilon^{\alpha\beta}S_\perp^\alpha q_\perp^\beta}
     {\vec{q}^2_\perp}
\sum\limits_{ab}
\int d^2k_{1\perp}d^2k_{2\perp}d^2\lambda_\perp
\nonumber \\
&&\times \frac{\vec{k}_{1\perp}\cdot \vec{q}_\perp}{M_P}\,
   x_a\, q_{Ta}^{\rm SIDIS}(x_a,k_{1\perp})\,
   x_b\, f_b^{\rm SIDIS}(x_b,k_{2\perp})
\label{x-ssa-tmd}\\
&&\times \left[S_{ab\to cd}(\lambda_\perp)\, H_{ab\to cd}^{\rm
Sivers}(P_\perp^2)\right]_c\,
\delta^{(2)}(\vec{k}_{1\perp}+\vec{k}_{2\perp}+
\vec{\lambda}_\perp-\vec{q}_\perp) \, , \nonumber
\end{eqnarray}
where apart from the functions already given, $f_b^{\rm SIDIS}$
denotes the unpolarized TMD quark distribution and $S_{ab\to cd}$
is the soft factor mentioned above \cite{JiQiuVogYua06,qvy-long}.
Because of the
color flow into the jets, the product of the soft and hard factors
will involve a sum over separate color amplitudes in the full
factorization formalism \cite{Botts:1989kf,{Kidonakis:1997gm}},
which has been represented by a trace $[~~]_c$ in color
space in the above equation. Our calculation of initial-state
collinear gluon radiation described above would not be sensitive to
this complexity of the color flow, but we emphasize that the
definition of the parton distributions cannot be affected by
it \cite{Kidonakis:1997gm,qvy-long}.

In Eq.~(\ref{x-ssa-tmd}), we have chosen TMD parton distributions
defined in SIDIS because of the dominance of final-state
interactions. Choosing TMD parton distributions defined according
to the Drell-Yan process would change the sign of the partonic
hard factors, but not affect the overall sign of the physical
cross section. Based on our explicit calculation here and the
generalized factorization property of the Drell-Yan process at low
$q_\perp$ \cite{Collins:1984kg,JiMaYu04}, and because of the
similarity in kinematics between two processes, we expect the generalized
factorization formula in Eq.~(\ref{x-ssa-tmd}) to be valid for
describing the single-transverse-spin dependent cross section for
the di-jet momentum imbalance in hadronic collisions in the
kinematic region where $P_\perp\gg q_\perp \gtrsim \Lambda_{\rm
QCD}$.

A key feature of the factorization is that the perturbatively
calculated short-distance hard factors should not be sensitive to
details of the {\it factorized} long distance physics.  We tried,
as a test, to derive all short-distance hard factors by using this
factorization formula and the Brodsky-Hwang-Schmidt model for SSAs
\cite{BroHwaSch02}. We were able to recover all hard-scattering
factors in this way. We note that the same hard factors $H_{ab\to
cd}^{\rm Sivers}$ as above were also found for the weighted
(integrated) SSA for di-jet production in \cite{mulders}. A
detailed comparison between the approach of \cite{mulders} and
ours will be presented in Ref.~\cite{qvy-long}. An all-order proof
of the above factorization formula, if correct, is still needed
and is beyond the scope of this paper.

{\bf 4. Summary.} We have studied the single-transverse-spin
dependent cross section $\Delta\sigma(S_\perp)$ for di-jet
production momentum imbalance in high-energy hadronic collisions
in a kinematic region where $P_\perp\gg q_\perp\gg \Lambda_{\rm
QCD}$, and calculated contributions from both initial- and
final-state interactions. At the leading order in
$q_\perp/P_\perp$, the $q_\perp$ and $P_\perp$ dependences in our
calculated results are decoupled and can be factorized into a
single-scale hard factor that depends on $P_\perp$, and into
perturbatively generated TMD parton distributions. This
factorization occurs for each partonic channel. Overall,
final-state interactions turn out to give the dominant
contribution to $\Delta\sigma(S_\perp)$  \cite{qvy-long}. We
therefore expect the SSA in di-jet production to have the same
sign as the Sivers asymmetry in SIDIS.

We have found that our results are consistent with a more general
TMD factorization formula, given in Eq.~(\ref{x-ssa-tmd}), which
we propose to be the correct approach to describing the
single-transverse spin asymmetry in di-jet production at hadron
colliders when $P_\perp\gg q_\perp$. We emphasize that obviously
the result of our first-order calculation is not able to actually
prove this factorization, but should rather be regarded as a
``necessary condition'' for such a factorization to hold. A full
proof remains an important challenge for future work. Also, we
recall that we have limited ourselves to the case of collinear
initial-state gluon radiation. The effects of large-angle
soft-gluon emission have been neglected, even though we have
indicated their likely role in Eq.~(\ref{x-ssa-tmd}). A proof of
TMD factorization in this process would naturally incorporate a
study of this soft factor.

If the proposed factorization formula is valid,
the di-jet momentum imbalance at RHIC could be described
by the same TMD parton distributions as those used to describe the
SIDIS and Drell-Yan processes. Because both initial- and final-state
interactions are present, the di-jet momentum imbalance is sensitive
to {\it different} short distance dynamics, and it will
be an excellent process to test QCD factorization and the
universality of the TMD parton distributions. In addition it should
give valuable information on the partons' transverse motion in the nucleon.


\begin{acknowledgments}
We thank C.J. Bomhof, S.J. Brodsky, J.C. Collins, X. Ji, A. Metz,
P.J. Mulders, and G. Sterman for useful discussions. We are
grateful to RIKEN, Brookhaven National Laboratory and the U.S.
Department of Energy (grant number DE-FG02-87ER40371 and contract
number DE-AC02-98CH10886) for providing the facilities essential
for the completion of this work. J.Q. thanks high energy theory
group at Argonne National Laboratory for its hospitality during
the writing of this work.
\end{acknowledgments}



\begin{thebibliography}
\frenchspacing

\bibitem{Bunce}
  G.~Bunce {\it et al.},
  Phys.\ Rev.\ Lett.\  {\bf 36}, 1113 (1976).

\bibitem{E704} see, for example:
  D.~L.~Adams {\it et al.},
  Phys.\ Lett.\ B {\bf 261}, 201 (1991);
  Phys.\ Lett.\ B {\bf 264}, 462 (1991);
K.~Krueger {\it et al.}, Phys.\ Lett.\ B {\bf 459}, 412 (1999).

\bibitem{rhic}
  J.~Adams {\it et al.}, 
  Phys.\ Rev.\ Lett.\  {\bf 92}, 171801 (2004);
%
  S.~S.~Adler, 
Phys.\ Rev.\ Lett.\  {\bf 95}, 202001 (2005);
%
  F.~Videbaek, 
AIP Conf.\ Proc.\  {\bf 792}, 993 (2005).

\bibitem{hermes}
A.~Airapetian {\it et al.}, 
Phys.\ Rev.\ Lett.\  {\bf 84}, 4047 (2000);
  {\it ibid}.,  {\bf 94}, 012002 (2005);
  V.~Y.~Alexakhin {\it et al.}, 
  Phys.\ Rev.\ Lett.\  {\bf 94}, 202002 (2005).


\bibitem{review} for reviews, see:
M.~Anselmino, A.~Efremov and E.~Leader,
Phys.\ Rept.\  {\bf 261}, 1 (1995) [Erratum-ibid.\  {\bf 281}, 399
(1997)];
  Z.~t.~Liang and C.~Boros,
  Int.\ J.\ Mod.\ Phys.\ A {\bf 15}, 927 (2000);
V.~Barone, A.~Drago and P.~G.~Ratcliffe,
Phys.\ Rept.\  {\bf 359}, 1 (2002).

\bibitem{et}
  A.~V.~Efremov and O.~V.~Teryaev,
  Sov.\ J.\ Nucl.\ Phys.\  {\bf 36}, 140 (1982)
  [Yad.\ Fiz.\  {\bf 36}, 242 (1982)];
  Phys.\ Lett.\ B {\bf 150}, 383 (1985).

\bibitem{qs-ssa}
J.~W.~Qiu and G.~Sterman,
Phys.\ Rev.\ Lett.\  {\bf 67}, 2264 (1991);
  Nucl.\ Phys.\ B {\bf 378}, 52 (1992);
Phys.\ Rev.\ D {\bf 59}, 014004 (1999).

\bibitem{JiQiuVogYua06}
  X.~Ji, J.~W.~Qiu, W.~Vogelsang and F.~Yuan,
Phys.\ Rev.\ Lett.\ {\bf 97}, 082002 (2006);
  Phys.\ Rev.\ D {\bf 73}, 094017 (2006);
  Phys.\ Lett.\ B {\bf 638}, 178 (2006).


\bibitem{BoeVog03}
  D.~Boer and W.~Vogelsang,
  Phys.\ Rev.\ D {\bf 69}, 094025 (2004).

\bibitem{mulders} A.~Bacchetta, C.~J.~Bomhof, P.~J.~Mulders and F.~Pijlman,
  Phys.\ Rev.\ D {\bf 72}, 034030 (2005);
  C.~J.~Bomhof and P.~J.~Mulders,
  JHEP {\bf 0702}, 029 (2007).

\bibitem{Bomhof:2007su}
  W.~Vogelsang and F.~Yuan,
  Phys.\ Rev.\ D {\bf 72}, 054028 (2005);
  C.~J.~Bomhof, P.~J.~Mulders, W.~Vogelsang and F.~Yuan,
  arXiv:hep-ph/0701277.

\bibitem{star-dijet1}
 J.~Balewski, talk presented at the SPIN 2006 Symposium,
Kyoto, Japan, October 2-7, 2006, arXiv:hep-ex/0612036.

\bibitem{qs-ssa-fac}
  J.~W.~Qiu and G.~Sterman,
  AIP Conf.\ Proc.\  {\bf 223}, 249 (1991);
  Nucl.\ Phys.\  B {\bf 353}, 137 (1991).

\bibitem{qvy-long}
J.~W.~Qiu, W.~Vogelsang, and F.~Yuan, in preparation.

\bibitem{Kouvaris:2006zy}
  C.~Kouvaris, J.~W.~Qiu, W.~Vogelsang and F.~Yuan,
  Phys.\ Rev.\  D {\bf 74}, 114013 (2006).


\bibitem{css-fac}
  J.~C.~Collins, D.~E.~Soper and G.~Sterman,
  Adv.\ Ser.\ Direct.\ High Energy Phys.\  {\bf 5}, 1 (1988)
  and references therein.


\bibitem{Collins:1984kg}
  J.~C.~Collins, D.~E.~Soper and G.~Sterman,
  Nucl.\ Phys.\  B {\bf 250}, 199 (1985).

\bibitem{JiMaYu04}
  X.~Ji, J.~P.~Ma and F.~Yuan,
  Phys.\ Rev.\ D {\bf 71}, 034005 (2005);
Phys.\ Lett.\ B {\bf 597}, 299 (2004).

\bibitem{Siv90}
D.~W.~Sivers,
Phys.\ Rev.\ D {\bf 41}, 83 (1990);
Phys.\ Rev.\ D {\bf 43}, 261 (1991).

\bibitem{BroHwaSch02}
S.~J.~Brodsky, D.~S.~Hwang and I.~Schmidt,
Phys.\ Lett.\ B {\bf 530}, 99 (2002);
Nucl.\ Phys.\ B {\bf 642}, 344 (2002).

\bibitem{Col02}
J.~C.~Collins,
Phys.\ Lett.\ B {\bf 536}, 43 (2002).

\bibitem{BelJiYua02}
X.~Ji and F.~Yuan,
Phys.\ Lett.\ B {\bf 543}, 66 (2002);
A.~V.~Belitsky, X.~Ji and F.~Yuan,
Nucl.\ Phys.\ B {\bf 656}, 165 (2003).


\bibitem{BoeMulPij03}
D.~Boer, P.~J.~Mulders and F.~Pijlman,
Nucl.\ Phys.\ B {\bf 667}, 201 (2003).

\bibitem{Botts:1989kf}
  J.~Botts and G.~Sterman,
  Nucl.\ Phys.\  B {\bf 325}, 62 (1989).

\bibitem{Kidonakis:1997gm}
  N.~Kidonakis and G.~Sterman,
  Nucl.\ Phys.\  B {\bf 505}, 321 (1997);
 N.~Kidonakis, G.~Oderda and G.~Sterman,
Nucl.\ Phys.\  B {\bf 531}, 365 (1998).
\end{thebibliography}
\end{document}